\documentclass[11pt,a4paper]{article}
\usepackage[nohead, nomarginpar, margin=1in, foot=.25in]{geometry}

\usepackage{graphicx}
\usepackage{color}
\usepackage{xcolor}
\usepackage{amsmath}
\usepackage{amsmath,bbm}

\usepackage{amssymb,bm}

\usepackage{csquotes}
\usepackage{subcaption}

\usepackage[T1]{fontenc}
\usepackage[utf8]{inputenc}
\usepackage{authblk}

\usepackage{soul}
\usepackage[LGRgreek]{mathastext}

\usepackage{url}

\usepackage{lineno}

\usepackage{cancel} 

\DeclareSymbolFont{greekletters}{OML}{cmr}{m}{it}
\DeclareMathSymbol{\varrho}{\mathalpha}{greekletters}{"25}

\usepackage{cite}
\date{}

\def\@listcomma@comma{\@ifnum{\@tempcnta>\tw@}{,}{}}

\begin{document}

\title{Centrality dependence of nuclear suppression of D mesons in $p$+Pb collisions at $\sqrt{s_{NN}}$ = 5.02 TeV}
\author[1]{Srikanta Kumar Tripathy}
\author[2, \footnote{younus.presi@gmail.com}]{Mohammed Younus}
\author[3, \footnote{sudipan86@gmail.com}]{Sudipan De}

\affil[1]{Department of Atomic Physics, Eotvos Lorand University, Budapest, H-1117, Hungary}
\affil[2] {Department of Physics, Nelson Mandela University, Port Elizabeth, 6031, South Africa }
\affil[3] {Department of Physics, Dinabandhu Mahavidyalaya (Bongaon), North 24 Parganas, PIN - 743235, West Bengal, India}

\maketitle

\begin{abstract}
\end{abstract}

In this paper we have shown theoretical model comparisons with ALICE results for average D mesons (D$^0$, D$^+$, D$^{*+}$ and $D^+_s$) in $p$+Pb collisions at $\sqrt{s_{NN}}$ = 5.02 TeV for various centralities. Transport calculations of AMPT and calculations from heavy quark pQCD model, NLO(MNR) have been used for the study of $p_T$ dependent nuclear modification factors in terms of collision centrality ($\displaystyle{Q_{pPb}}$) and the central-to-peripheral ratios ($\displaystyle{Q_{cp}}$) of D mesons.
It is found that NLO model with its parametrized $\displaystyle{k_T}$ broadening scheme produces results those closely match with the published D-meson data of $p$+Pb collisions from ALICE. Likewise AMPT transport calculation shows a strong centrality dependence in results but underestimates the experimental data. The differences of both models with experimental data have been discussed.

\noindent\\
\textit{Keywords: $p$+$p$ and $p$+Pb collision; D-meson; Cold-Nuclear matter effect.}\\
\textit{PACS Nos.:14.40.Lb,14.65.Dw,25.75.-q} \\

\section{Introduction}
\label{intro}

A deconfined phase of quark and gluons commonly known as quark gluon plasma (QGP) is formed when heavy ions collide at relativistic speeds \cite{qgp1a,qgp1b,qgp2a,qgp2b}. This novel matter survives for a infinitesimally small amount of time ($\sim$10$^{-23}$ seconds) under extreme conditions and therefore couldn't be observed directly. Only signals those originate from the matter itself might survive and measured after the freeze-outs.~\cite{qgprev1a,qgprev1b,qgprev2}. Data available from the Large Hadron Collider (LHC) experiments at CERN has opened ways for extracting information on quark gluon plasma while phenomenological models are able to explain this data and  physics behind it. These analyses have prompted precise measurements at ALICE at CERN and CBM at FAIR and would help us improve our models.

Quenching of jets or high momenta particles is one of the most significant feature of QGP. Calculation of nulcear modification factor ratio, $R_{AA}$ shows that the momentum spectra of hadrons in heavy ion collisions are suppressed compared to scaled hadron momentum spectra from $p$+$p$ collisions. \cite{suppress1A,suppress1B,suppress1C, suppress2,suppress3,suppress4}. This 
suggests a quenching effect within a hot and dense partonic matter. A small suppression is also observed in hadronic phase. Similar suppression effects have been observed for 
high $p_T$ heavy quarks as observed from D or
B mesons data which is of same order as the light partons \cite{exp1,exp2,sd1,sd2,Abelev1}. The reason for such suppression irrespective of particle type is assumingly due to the absence of nuclear effects in  $p$+$p$ collisions. Therefore it is possible to use $p$+$p$ collision as a baseline and scale it to $p$+Pb or Pb+Pb data by a factor only. Heavy quarks which are formed in the pre-equilibrium phase of heavy ion collisions, carry the information on QGP through their suppression. Moreover, similar to jets and photons, heavy quarks are also affected by the collective medium flow. Some recent results have shown that contrary to popular notion that $p+p$ collisions might also produce quenching or flow like effects particularly at very high c.m. energy or high multiplicity region ~\cite{Deng:2011at}. Thus it is important to discern any effects nuclear matter might bring in and can be distinguished from small system. Whether similar observation is also present in $p$+Pb collisions which is somewhat in between $p$+$p$ and Pb+Pb systems needs to be studied properly. It is known that particle spectra is also modified by initial cold nuclear matter (CNM) \cite{cnm1,cnm2} even before the formation of QGP or any partonic system. This also makes heavy ion collisions different from proton on proton collisions. However CNM effects are easily masked by QGP effects  except at very low momentum region. The contributions  of cold nuclear matter effects can be separated from all other effects due to QGP by phenomenological models and specific experiments those could be directed to such purposes. Hadron-Ion collisions, specifically $d$+ Au and $p$+Pb collisions could give us ground to study these initial nuclear effects on not only particle flow but also on its quenching, momentum and azimuthal correlations etc.~\cite{Vogt:2019xmm,Younus:2017xmg}. The source for pre-QGP nuclear effects lies in the fact that protons inside the nucleus are correlated and this gives rise to effects such shadowing-anti-shadowing phenomena, multiple nucleon and partonic interaction, iso-spin effects etc to name a few. At top collider energies, these effects alter the high gluon density within the nucleus. This may very well modify our understanding of hot and dense QGP.
Shadowing-anti-shadowing is mathematically wriiten as , $R_s \equiv F_A(x,Q^2)/(A*F_p(x,Q^2))$. It is important to understand that at both low and high parton momentum fraction, $x$ sea gluons distribution is affected by the correlated protons within finite nucleus and modify the parton distribution functions (PDFs).~\cite{shadow1a,shadow1b} This phenomenon is one of the interesting features of cold nuclear matter. Similarly, a second phenomenon that affects the final particle spectra is multiple re-scattering of the interacting nucleons and their partons. This effect is also called Cronin 
effect \cite{cronin1} and is responsible for momentum broadening of the parton distributions within the nucleus. This particular feature had been observed at the RHIC energy for non-photonic electrons' data, and also shows an enhancement in the spectra at $p_T < 4.0 $ GeV \cite{phenix1}. 
Some current model studies suggest that this particular effect is observed 
in the low and mid-p$_T$ regions and may not be much effective in higher side of the momentum. We will discuss more of this later. In any case for a $p$+Pb scenario even if a small system of hot and dense matter is formed similar to high multiplicity $p$+$p$ collisions, CNM effects might be overwhelmingly visible. Also the collisions of protons with Lead ions at various centralities will also bring out difference of $p$+Pb system with $p$+$p$ system at the most peripheral and with Pb+Pb system at the most central collision scenarios.

As shown in earlier literatures heavy quark pair is produced mostly in pre-equilibrium stage of heavy ion collisions~\cite{charm1a,charm1b,charm1c}. It is also 
known that heavy quarks remain free to 
probe QGP without altering much of the effects due to cold nuclear matter. 
Some results of from $p$+Pb data and earlier $d$+Au data~\cite{charm2a,charm2b,charm2c} on charm production, 
the value of $R_{pA}$ deviates from unity by almost 
15$\%$ in low and mid-p$_T$ regions, and even overshadowed the expected QGP effects in the said regions. A considerable effect due to cold nuclear matter on heavy 
quark production \cite{charm3} could be a possible reason behind it. It is known that particle suppression has strong dependence on collision centrality with particle multiplicity and QGP freeze-out times depending on non-centrality of the collisions. Similarly it can be assumed different centralities in collision geometry may have impact on particle production in case of $p$+Pb collisions. We will return to this in subsequent sections. The current work aims to highlight the effects of CNMs on heavy quark quenching or modification factor at various centralities of $p$+Pb collisions.

This paper is organised as follows. In the section \ref{Models_used} the transport model of AMPT and the pQCD model of NLO-MNR have been employed to study D-meson suppression factors $\displaystyle{Q_{pPb}}$ and $\displaystyle{Q_{cp}}$ in $p$+Pb collisions at various collision centralities. The calculations have been done at $\sqrt{s_{NN}}$ = 5.02 TeV. 
In the section \ref{Results_and_discussion} we discuss our results from these models and compare them with ALICE data. 
Then comes the summary section \ref{Summary} followed by the bibliography.


\section{Models used}
\label{Models_used}

\subsection{The AMPT model}
\label{Ampt}

\noindent We have used string melting version of A Multiphase Transport Model (AMPT) \cite{ampt_model}  (version 26t5). This model uses HIJING (Heavy-Ion Jet Interaction Generator)\cite{hij} for spatial and momentum distribution of strings and minijet partons.\\ 
Eikonal formalism is used to deal with scattering among nucleons, distribution of which are Wood-Saxon in profile. Production of minijet happens if momentum transfer ($Q^{2}$) is greater than some cut off momentum ($p_{0}$), while the opposite ($Q^{2} < p_{0} $) leads to production of strings. Depending on spin and flavor of valence quarks, these produced minijet and strings  get converted into partons. Partons which have minimum distance conditions ($\leq \sqrt{\sigma/\pi}, \ \sigma$ is the cross section for partonic two-body scattering),  interaction between them dealt by Zhang's Parton Cascade (ZPC) model \cite{zpc}. ZPC uses Boltzmann equation, where the differential cross-section (leading order, two body scattering) is given as follow:
\begin{equation}
\frac{d\sigma_{gg}}{dt} \approx  \frac{9\pi \alpha_s^2}{2(t - \mu^2)^2},
\end{equation}
where, $\alpha_s$, t and  $\mu$ are strong coupling constant, standard Mandelstam variables for momentum transfer
and screening mass of partonic matter respectively.
A quark coalescence model is used to form baryons or mesons once these partons stop interacting.  A relativistic transport model (ART) \cite{art_1, art_2} deals with resultant hadron cascade, which includes elastic and inelastic scatterings.\\

\subsection{The NLO model}
\label{Nlo}

The next-to-leading order, NLO-pQCD(MNR)\cite{MNR1,MNR2} model has been used to calculate charm \textit{c$\bar{\textit c}$} and bottom \textit{b$\bar{\textit b}$} pairs 
cross-sections in $p$+$p$ for RHIC and LHC collider energies \cite{jamil}. The model has been used to calculate heavy quark pair correlation in azimuthal angle and rapidity for both protons and heavy ions collisions at LHC energies. In these works, no medium effects were considered in particle cross-section for Pb+Pb  collisions and the effects of various orders in invariant matrix have been studied ~\cite{younus}.
The model can be used to produce heavy quark differential cross-section and $p_T$ spectra and can also be utilised 
to study particle observables by incorporating hot and dense nuclear matter effects (\textit{viz.} Pb+Pb and Au+Au collisions) 
and cold nuclear matter effects (\textit{viz.} $p$+Pb and $d$+Au collisions). In an earlier work, the model has been used to produce D-meson spectra for $p$+$p$ collisions at $\sqrt{s}$ = 7 TeV and D mesons' $R_{pPb}$ (min. bias) 
for $p$+Pb collisions at $\sqrt{s}$ = 5.02 TeV.~\cite{Baral:2016zap} 
In the present work instead of minimum bias configuration, the calculations have been extended to various collision centralities in $p$+Pb collision at same collider energy and various initial cold nuclear effects~\cite{shadow2,ramona1} have been included. Let us now move to describe the model briefly:

The $p_T$ differential spectrum of heavy quarks produced in $p$+$p$ collisions is defined in general 
as \cite{younus,jamil}
\begin{equation}
{E_1}{E_2}\frac{d\sigma}{d^{3}p_1 d^{3}p_2} = \frac{d\sigma}{dy_1 dy_2 d^{2}p_{T_1} d^{2}p_{T_2}}\  ,
\label{ini11}
\end{equation}

where, $y_1$ and $y_2$ are the rapidities of heavy quark and anti-quark 
and {\bf\it p}$_{ \textit {\bf T}_i}$ are their transverse momenta.\\
In the above
\begin{equation}
\begin{split}
 \frac{d\sigma}{dy_1 dy_2 d^{2}p_{T_1} d^{2}p_{T_2}} = & \ 2 x_{a}x_{b}\sum_{ij} \bigg[ f^{(a)}_{i}(x_{a},Q^{2})f_{j}^{(b)}(x_{b},Q^{2}) \frac {d\hat{\sigma}_{ij}(\hat{s},\hat{t},\hat{u})} {d\hat{t}} \\
& +  f_{j}^{(a)}(x_{a},Q^{2})f_{i}^{(b)}(x_{b},Q^{2}) \frac{d\hat{\sigma}_{ij}(\hat{s},\hat{u},\hat{t})}{d\hat{t}} \bigg] /(1+\delta_{ij}) \ ,\\
\label{ini2}
\end{split}
\end{equation}

Here $x_{a} $ and $x_{b} $ are the fractions of the momenta carried by the partons 
from their interacting parent hadrons.

CTEQ6.6 structure function \cite{cteq66} as obtained from LHAPDF library for protons has been used and 
EPS09 \cite{eps09} shadowing parameterization has been added 
to incorporate the initial nuclear effects on the parton densities for lead, Pb ions.

The differential cross-section for partonic interactions, $d\hat{\sigma}_{ij}/d\hat{t}$ 
is given by
\begin{equation}
\frac{d\hat{\sigma}_{ij}(\hat{s},\hat{t},\hat{u})}{d\hat{t}} = \frac{\left|M\right|^{2}}{16\pi\hat{s}^{2}} ,
\label{dsdt}
\end{equation}
where, $\left|M\right|^{2}$ (See Ref.~\citenum{combridge}) is the invariant amplitude for various 
partonic sub-processes both for leading order (LO) and next-to-leading order (NLO) processes as follows: \\
The physical sub-processes included for the leading order, 
$\cal{O}$ $(\alpha_{s}^{2}) $ production of heavy quarks are
\begin{equation}
\begin{split}
 g+g & \rightarrow Q+\overline{Q} \ and  \\
 q+\bar{q} & \rightarrow Q+\overline{Q} \ . 
\end{split}
\end{equation}
At next-to-leading order, $\cal{O}$ $(\alpha_{s}^{3})$ sub-processes 
included are as follows
\begin{equation}
\begin{split}
g+g & \rightarrow Q+\overline{Q}+g \ , \\
q+\bar{q} & \rightarrow Q+\overline{Q}+g \ and \\
g+q(\bar{q}) & \rightarrow Q+\overline{Q}+q(\bar{q}) .
\end{split}
\end{equation}

Re-scattering of nucleons and their partons include multiple hard scattering (parton-parton or parton-nucleon) or multiple soft scattering (nucleon-nucleon) within the nuclear volume. 
This is also called Cronin effect \cite{cronin1,accardi1}. 
The re-scatterings may lead to momentum broadening of the 
interacting partons and change the final particle and parton densities. 
This would ultimately give rise to deviations of $R_{pPb}$ and $Q_{pPb}$ from unity and is a signature of cold nuclear matter effect. Its contribution apart from shadowing to the hadron spectra compared to QGP effects can be discerned with the precise state-of-the-art experiments at LHC-CERN. The Cronin effect may vanish at large 
transverse momentum region or high collider energies \cite{accardi2,sharma1,levai1}, but visible in the low and mid $p_T$ region.~\cite{accardi1,gyulassy1A,gyulassy1B} 

We can now discuss briefly the mechanism of multiple re-scattering or Cronin effect. It can also be termed as $\displaystyle{k_T}$ broadening effect.  
The parton density function can be defined as
\begin{equation}
f^{(a)}_{i}(x_{a},Q^{2},k_T^2)=f_{i}^{(a)}(x_{a},Q^{2}).g_{p/A}(k_T^2) \ , \\
\end{equation}

where, $g_{p/A}(k_T^2)\propto exp[-k_T^2/\pi\,.\langle k_T^2\rangle_{pp/pA}]$ and $\langle k_T^2\rangle_{pA}= \langle k_T^2\rangle_{pp} + \langle k_T^2\rangle_A$.\\

The net transverse momentum kick, 
$\langle k_T^2\rangle_{pA}$~\cite{accardi2,gyulassy1A,gyulassy1B},
 is obtained by adding $\langle k_T^2\rangle_A$ from multiple scattering to the intrinsic $\langle k_T^2\rangle_{pp}$. It can be stated that this assumption roughly extrapolate $p+p$ system to $p$+A collision system. We have also included the impact factor into the current model so as to distinguish the re-scattering phenomena for collisions at various centralities. The $\langle k_T^2\rangle_A$ can be assumed as 
\begin{equation}
 \langle k_T^2\rangle_A = \delta^{2} . N(b,\sqrt{s}) .\ln \bigg( 1+\frac{p_T^2}{\delta^2/c} \bigg)
\end{equation}
where the parameter $\delta^2$ is average squared momentum kick per scattering($\sim$ 1 GeV$^2$/c$^2$). The number of average re-scattering, $\displaystyle{N(b, \sqrt{s})}$ at a given centrality and collision energy is defined as

\begin{equation}
  N(b,\sqrt{s})=\begin{cases}
    T_{pA}(b).\sigma_{NN}-1, & \text{if $T_{pA}.\sigma\geq n$}.\\
    n, & \text{otherwise}.
  \end{cases}
\end{equation}

The nuclear overlap function $\displaystyle{T_{pA}}$ could calculated at different centralities using Glauber model simulation. $\sigma_{NN}$ is taken to be 69 mb at $\sqrt{s_{NN}}$ = 5.02 TeV. $\displaystyle{n}$ is the lower limit of re-scattering parameter and is taken to be 4. However multiple hard scattering $>$ 4 can dissociate nucleons and is known as {\it re-scattering saturation}~\cite{accardi1,accardi2}.
After implementing the re-scattering method, fragmentation mechanism is applied on the produced charm quarks both from $p$+A and $p$+$p$
collisions into D-mesons. Schematically,
the fragmentation can be shown as
\begin{equation}
E\frac{d^3\sigma}{d^3p}=E_Q\frac{d^3\sigma}{d^3p_Q}\otimes D(Q\rightarrow H_M),
\label{eq18}
\end{equation}
where, the fragmentation of the heavy quark $Q$ into the heavy-meson $H_M$ is
described by the function $D_D(z)$. We have assumed that distribution of
$D(z)$, w.r.t. $z$, where $z=p_D/p_c$, is used to calculate total $D$-mesons and
is given by
\begin{equation}
D^{(c)}_D(z)=\frac{n_D}{z[1-1/z-\epsilon_p/(1-z)]^2},
\label{eq19}
\end{equation}
where, $\epsilon_p$ is the Peterson parameter $\simeq \ $0.12 and
is taken from Ref.~\citenum{peterson1}.
The normalization condition satisfied by the fragmentation
function is
\begin{equation}
\int_0^1 \, dz \, D(z)=1 .
\label{eq20}
\end{equation}

\section{Results and discussion}
\label{Results_and_discussion}
It should be recalled that in the present calculation, the current version of AMPT uses the coalescence mechanism for hadronization while NLO model incorporates pQCD techniques and fragmentation mechanism for hadronization. This definitely brings in uncertainties between two models shown. However heavy quarks being massive particles, these two hadronization mechanisms can work closely in the mid-p$_T$ region where both of them are valid. While at high p$_T$ region fragmentation process is dominant, at low p$_T$ (p$_T$ $<$ 1 GeV) coalescence mechanisms contributes majorly~\cite{Cao:2013ita,Cao:2015hia}. Furthermore we are more interested in highlighting the CNM effects on particle distribution, any such effects due to hadronization mechanisms are presumably nullified in the ratios $Q_{pPb}$ and $Q_{cp}$ where same mechanism is present in both numerator and denominator of the ratios. However the scenario might not be the same in case of light quarks or for all momenta of the particle. Any elaborate study on hadronization mechanisms and its effects on freeze-out surfaces particularly in context of heavy mesons production would be referred in the future publications. 

\begin{figure}[h]
\centering
\includegraphics[scale = 0.5]{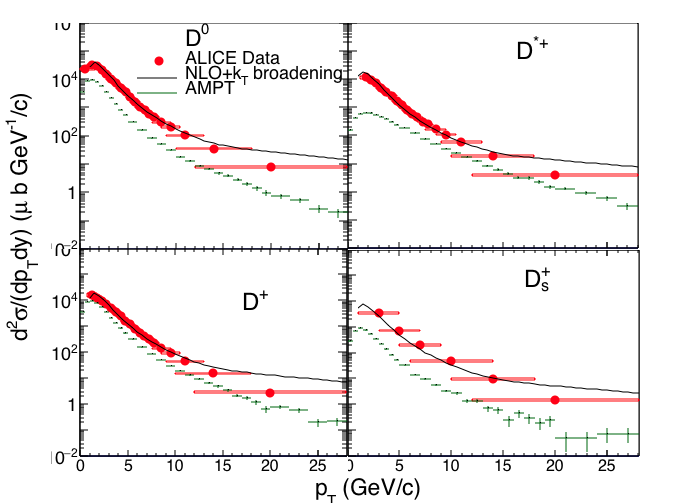}
\caption{\small(Color online) $p_T$ differential cross-section of D mesons in $p$+Pb at $\sqrt{s_{NN}}$ = 5.02 TeV. Solid markers represent the ALICE data points~\cite{alice_qcp_qppb}, while green marker represents AMPT calculations and black lines are from NLO with $k_T$ broadening calculations. The vertical lines in the AMPT calculation points represent the statistical uncertainties.}
\label{Fig:dSigdpT}
\end{figure}

In this article we have generated simulated events to compare with ALICE results on D-meson ($D^0$, $D^+$, $D^{*+}$ and $D_s^+$) in $p$+$p$ and $p$+Pb collisions~\cite{alice_qcp_qppb}. Mid-Rapidly ($|y_{cms}| <$ 0.5) used for $p$+$p$ system,  where as  -0.96 $<$ y$_{cms}$ $<$ 0.04 for $p$+Pb system. However, the p+p yield is corrected for the rapidity shift in $p$+Pb collisions. Scaling factors from Ref.~\citenum{tpp_tpPb}) are used to obtain cross-section. Also we make sure that,
 only direct production of D mesons have been considered and no B meson decay into D mesons is present.

We have shown $p_T$ differential cross-section of D mesons for p+Pb minimum bias collisions at $\sqrt{s_{NN}}$ = 5.02 TeV in Figure~\ref{Fig:dSigdpT}. Different panels show different species of D mesons such as   $D^0$, $D^+$, $D^{*+}$ and $D_s^+$. The solid circles show the experimental data points measured in ALICE~\cite{alice_qcp_qppb}. The lines show different model calculations. The solid black lines represent NLO calculations and green markers represent the calculations using AMPT. NLO shows a good agreement with data upto 15 GeV/c of $p_{T}$ while AMPT underestimates the data for all $p_{T}$ region for all D mesons. In case of NLO calculations, the next to leading order contributions start to dominate at high p$_T$ which leads to increased production of high momentum heavy quark pairs. A cut-off based on heavy quark mass has been included but an additional cut-off based on logarithmic resumption of diagrams at NLO level is required to curtail this over-production. Such resumption is present in FONLL model but it has its own limitation for not being able to calculate two particle correlation, azimuthal distribution etc. of heavy quarks which NLO pQCD does. Furthermore as mentioned earlier that while calculating nuclear modification factor, these effects may be canceled from both numerator and denominator. On the other hand AMPT has several inbuilt factors whose presence or absence lead to mismatch between data and the model. One of these effects is decay contribution to total cross-sections which is neglected in the present calculations and may have considerable effects on the outcome. Also there is an absence of next-to-leading order contribution to particle generation in AMPT and an additional presence of momentum and energy loss due to multiple scattering of partons both in initial and latter stages nuclear medium within the transport model. Combination of these factors contributes to the decrease in production cross-section of particle for both $p$+Pb and Pb+Pb collisions. Hence AMPT underestimates the data by a certain factor while the shapes of both model and data are similar.

\begin{figure}[h]
\centering
\includegraphics[scale = 0.25]{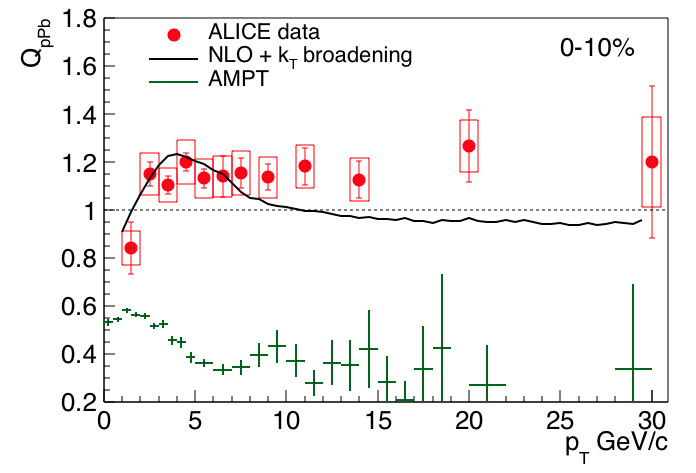}
\includegraphics[scale = 0.25]{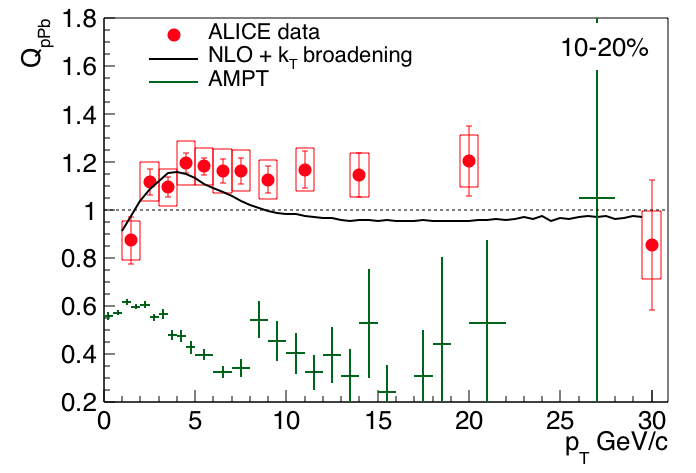}
\includegraphics[scale = 0.25]{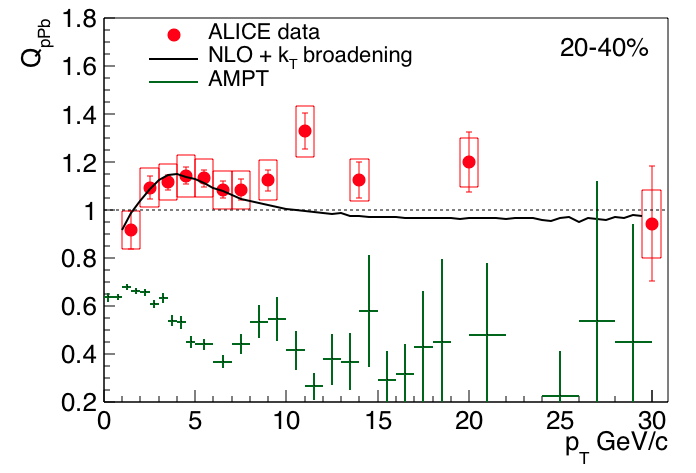}
\includegraphics[scale = 0.25]{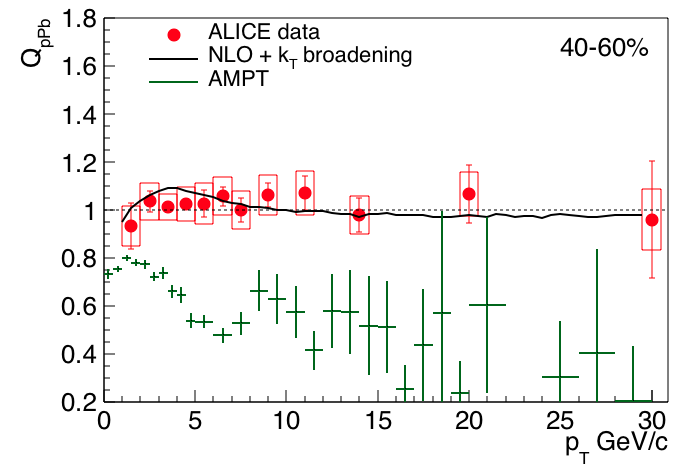}
\includegraphics[scale = 0.25]{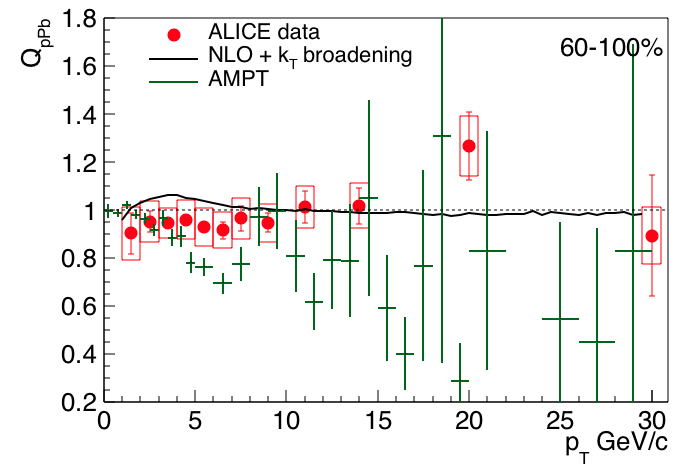}
\caption{\small(Color online) $Q_{pPb}$ of average D-meson ($D^0$, $D^+$, $D^{*+}$ and $D_s^+$) in $p$+Pb at $\sqrt{s_{NN}}$ = 5.02 TeV. Solid markers represent the ALICE data points~\cite{alice_qcp_qppb}, while green marker represents AMPT calculations and black lines are from NLO with $k_T$ broadening. Different panels represent different centrality classes.}
\label{Fig:QpPb}
\end{figure}

ALICE has measured the nuclear modification factor ($Q_{pPb}$) in various centrality intervals~\cite{alice_qcp_qppb}. $Q_{pPb}$ can be defined as follows:
\begin{align}
Q_{pPb} = \frac{(d^2N/dp_{T}dy)^i_{pPb} }{T^i_{pPb}(b)\times(d^2\sigma/dp_{T}dy)_{pp}} 
\end{align}
where, $(d^2N/dp_{T}dy)^i_{pPb}$ is the yield of D mesons in $p$+Pb collisions in different centrality classes, ($d^2\sigma/dp_{T}dy)_{pp}$ is the cross-section measured in p+p collisions at same center of mass energy and $\displaystyle{T_{pPb}^i(b)}$ is nuclear overlap function calculated at a particular centrality or impact parameter `$\displaystyle{b}$'. `i' stands for various centrality classes. Here we have used $p$+$p$ collisions as baseline at $\sqrt{s_{NN}}$= 5.02 TeV.

\begin{figure}[h]
\centering
\includegraphics[scale = 0.25]{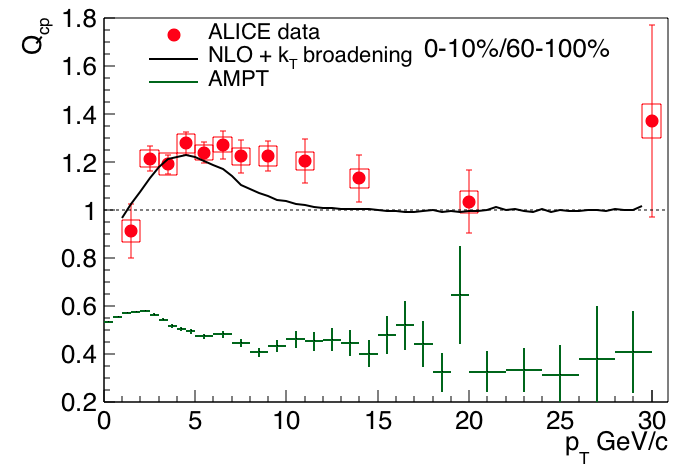}
\includegraphics[scale = 0.25]{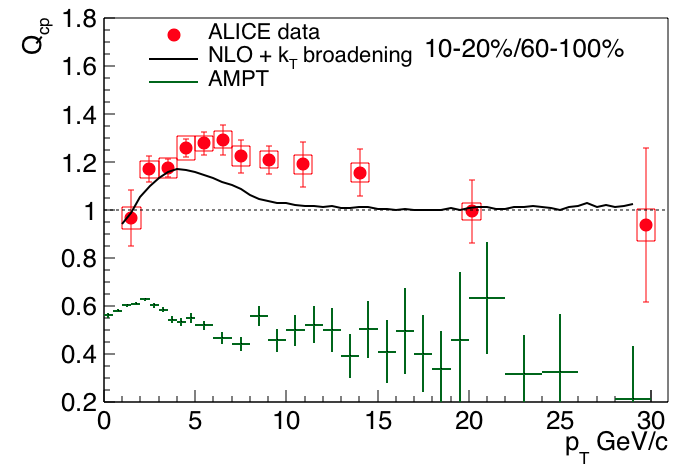}
\includegraphics[scale = 0.25]{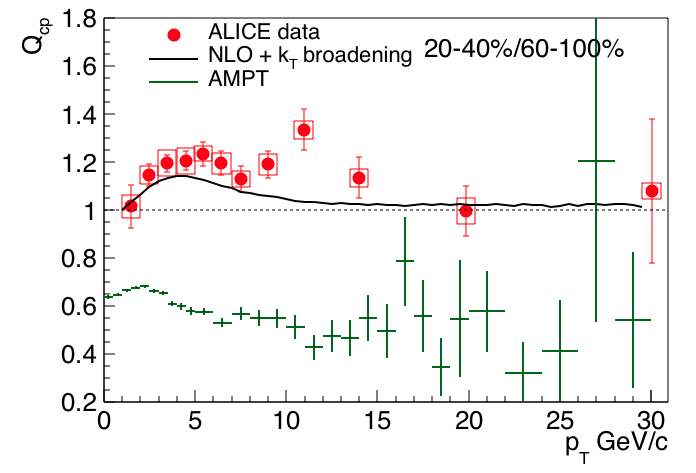}
\includegraphics[scale = 0.25]{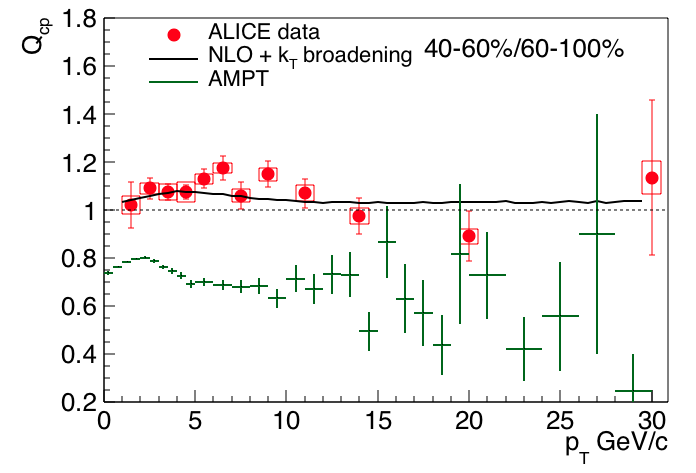}
\caption{\small(Color online) $Q_{cp}$ of avg. D-meson ($D^0$, $D^+$, $D^{*+}$ and $D_s^+$) in $p$+Pb at $\sqrt{s_{NN}}$ = 5.02 TeV. Solid markers represent the ALICE data points~\cite{alice_qcp_qppb}, while green marker represents AMPT calculations and black lines are from NLO with $k_T$ broadening. Different panels represent different centrality classes.}
\label{Fig:Qcp}
\end{figure}

Figure~\ref{Fig:QpPb} shows $\displaystyle{Q_{pPb}}$ of D mesons for several centrality classes such as 0-10\%, 10-20\%, 20-40\%, 40-60\% and 60-100\%. The solid symbols show the ALICE experimental results and the solid lines and markers represent NLO calculations and AMPT expectations respectively. The vertical lines in the data points represent statistical uncertainties and the boxes show systematic uncertainties. Data show a hint of enhancement at 2 $< p_{T} <$ 6 GeV/c for 0-40\% centrality. NLO calculations show similar enhancement and explain the data well upto 10 GeV/c $p_{T}$. For peripheral collisions (40-100\%) NLO explain the data very well within the uncertainties. 
Although NLO have nuclear shadowing feature and momentum broadening effect (Cronin) due to re-scattering, here momentum broadening effect is seemingly the dominant one.
 The results from AMPT with its shadowing and nuclear matter multiple scattering (scattering energy loss) under-estimate the experimental data. This shows the prominent contribution of initial cold nuclear matter (CNM) effects and multi-parton scattering effects, for the entire $p_{T}$ range in this model.

Recently ALICE has measured a new observable known as $Q_{cp}$~\cite{alice_qcp_qppb}. 
This may be defined as:\\
\begin{align}
Q_{cp} = \frac{ (d^2N/dp_{T}dy)^i_{pPb} /T^i_{pPb}(b)} { (d^2N/dp_{T}dy)_{pPb}^{60-100\%}/T_{pPb}^{60-100\%}(b)} 
 \end{align}

$\displaystyle{Q_{cp}}$ is independent of p+p cross-section and the spectra from most peripheral collisions (60-100\%) is used as reference. This reduces uncertainties coming from p+p measurements and we can get a more clear picture. Figure~\ref{Fig:Qcp} shows $\displaystyle{Q_{cp}}$ of average D mesons for four different centralities. A significant rise of $\displaystyle{Q_{cp}}$ is observed in central collisions (0-40\%) within 3 $< p_{T} <$ 7 GeV/c. For the first time we are trying to understand this experemnatl observation by comparing with different models. NLO + $k_T$ broadening results show closer affinity to ALICE results. However the shape of the results differ considerably after 6 GeV in momentum. Also the re-scattering effect has considerable effect at low momenta and even seems to overcome shadowing effect to some extent. This is corroborated by the fact that although D mesons nuclear factor shows a dip at low momenta showing shadowing effect, there is also a raise above unity. The magnitude of the factor calculated from NLO do not match the data entirely and it suggests that other factors such as energy loss etc. due to CNM might have a role to play. The modification factor is almost flat at high momenta with small discernible differences between centralities. It is also seen from the figures that NLO model with CNMs have very small effects at high momenta region and with the data having large errors at the end regions, it would be difficult to study the effects of CNMs in NLO model for high momenta particles. However the results do show effects of centralities on CNM effects in the intermediate momentum region. On the other hand the transport model AMPT under-predict the magnitude and shape of the experimental results for all centralities and for entire $p_{T}$ region.

\begin{figure}[h]
\centering
\includegraphics[scale = 0.3]{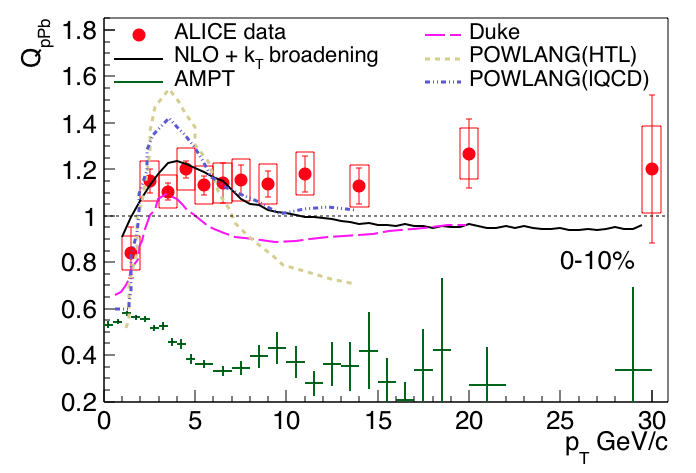}
\caption{\small(Color online) $Q_{pPb}$ of D-meson in $p$+Pb at $\sqrt{s_{NN}}$ = 5.02 TeV for 0-10\% centrality. Solid markers represent the ALICE data points~\cite{alice_qcp_qppb}, theoretical curves show the calculations from POWLANG(HTL) (0-20\%), POWLANG(lQCD) (0-20\%) and Duke  (0-10\%), which are also from~\cite{alice_qcp_qppb}. Green markers represent AMPT calculation and black line is from NLO with $k_T$ broadening calculation.}
\label{Fig:QpPbsum}
\end{figure}

Figure~\ref{Fig:QpPbsum} shows $\displaystyle{Q_{pPb}}$ as a function of $p_{T}$ of average D mesons for most central (0-10\%) collisions. Here we compare our results with other model expectations described in the Ref.~\citenum{alice_qcp_qppb}. All these models describe the data well upto 3-4 GeV/c of $p_{T}$ and can not reproduce the trend of the data at higher $p_{T}$. Whereas, NLO with momentum broadening reproduce the data upto around 10 GeV/c of $p_{T}$ and shows similar trend as data at high $p_{T}$ (although the error bars in data are too large at high p$_T$ to make any conclusion). The Duke~\cite{Xu:2015iha} and POWLANG~\cite{Beraudo:2015wsd}  both are transport models with assumption of QGP formation in $p$+Pb collisions.
Whereas NLO with momentum broadening is a theoretical models that include only CNM effects.This suggests that data is better described by the model which include initial state effect rather than final state effect in $p$+Pb collisions.

\section{Summary} 
\label{Summary}
\noindent
We have investigated D-meson nuclear modification factor in $p$+Pb collisions at $\sqrt{s_{NN}}$ = 5.02 TeV using simulation models like NLO and AMPT. Our results are compared with published ALICE results \cite{alice_qcp_qppb}.

We observe that NLO with momentum broadening describes the data much better than that of transport model AMPT.
AMPT results underestimates $Q_{pPb}$ and $Q_{cp}$ experimental data, although this model incorporates shadowing effect. While NLO models (which also incorporates shadowing effect) showed CNM effects at low and intermediate p$_T$s and much closer to experimental observations, the CNM effects and its centrality dependencies are under-whelmingly indiscernible at high p$_T$ regions.
So we may conclude that magnitude of $Q_{pPb}$ and $Q_{cp}$ in AMPT due to its additional partonic and hadronic transport parts have considerable effects on particle production. 
We deduce, additional mechanism in needed to be incorporated in AMPT, to explain its production cross-section for resonance particle viz., D$^{*+}$.
 Experimental results are better described by initial state model NLO with momentum broadening rather than the transport model AMPT, POWLANG and Duke suggesting domination of initial CNM effect than that of final state effect.

As concluding remarks, since $Q_{pPb}$ and $Q_{cp}$ in our calculations deviates from unity at low and mid-p$_T$ for all centralities, the initial cold nuclear matter effects incorporated in the models and their centrality dependencies play very important roles in describing the nuclear matter effects on heavy quark production in both heavy ions and hadron-ion collisions.

\section*{Acknowledgments}

SKT acknowledges support from NKFIH financial grant FK-123842 for this study. SD acknowledges the financial support by 
DST-INSPIRE program of Government of India.

\end{document}